\documentclass[twocolumn,amsmath,amssymb,prl,10pt,nofootinbib,superscriptaddress]{revtex4}
\usepackage{ graphicx, float,amsmath, amsmath, amssymb}
\usepackage[export]{adjustbox}

\def\be{\begin{equation}}
\def\ee{\end{equation}}
\def\bea{\begin{eqnarray}}
\def\eea{\end{eqnarray}}
\def\bse{\begin{subequations}}
\def\ese{\end{subequations}}

\usepackage[breaklinks, colorlinks, citecolor=blue]{hyperref}
\usepackage[labelsep=period]{caption}
\begin{document}
\title{Evaporation Spectrum of Black Holes from a Local Quantum Gravity Perspective}
\author{Aur\'elien Barrau}%
\affiliation{%
Laboratoire de Physique Subatomique et de Cosmologie, Universit\'e Grenoble-Alpes, CNRS/IN2P3\\
53, avenue des Martyrs, 38026 Grenoble cedex, France
}
\date{\today}
\begin{abstract} 
We revisit the hypothesis of a possible line structure in the Hawking evaporation spectrum of black holes. Because of nonperturbative quantum gravity effects, this would take place arbitrarily far away from the Planck mass. We show, based on a speculative but consistent hypothesis, that this naive prediction might in fact hold in the specific context of loop quantum gravity. A small departure from the ideal case is expected for some low-spin transitions and could allow us to distinguish several quantum gravity models. We also show that the effect is not washed out by the dynamics of the process,  by existence of a mass spectrum up to a given width, or by the secondary component induced by the decay of neutral pions emitted during the time-integrated evaporation.
 \end{abstract}
\maketitle
\section{Introduction and model}
The Hawking radiation of black holes \cite{Hawking:1974sw}, characterized by a Planck law slightly modulated by grey-body factors  \cite{Page:1976df},
is one of the most robust predictions of quantum field theory on a curved background. It is also the perfect phenomenon to investigate possible deviations from the semiclassical dynamics due to nonperturbative quantum gravity effects.

Bekenstein and Mukhanov \cite{Bekenstein:1995ju} have suggested that an interesting way to account for quantum gravity at the effective level could be to assume that the area of a black hole (BH) can only take values proportional to a fundamental area assumed to be of the order of the Planck area. This has interesting consequences. In particular, this leads to the appearance of emission lines instead of a continuous spectrum, as we will explain in more details later. However, it was then shown that in the specific setting of loop quantum gravity (LQG) the use of the actual eigenstates of the area operator does {\it not} lead to a Bekestein-Mukhanov-like spectrum \cite{Barreira:1996dt}. The density of energy levels instead reads as $\rho(M)\sim \rm{exp}(M\sqrt{4\pi G/3})$ which means that the spectral lines are virtually dense in frequency for large masses. The phenomenology of evaporating black holes in LQG has, therefore, focused so far on the very last stages of the emission where lines are  distinguishable. This is theoretically interesting but  probably out of reach of any reasonable phenomenological approach.\\

We would like to revisit this conclusion, somewhat in the line of \cite{Yoon:2012cq}. Basically, what was assumed in virtually all LQG studies (see, {\it e.g.}, \cite{Barrau:2011md,Barrau:2015ana}) on evaporating black holes is that ``something" independent from quantum gravity triggers the emission of a particle from the black hole. The exact energy of this particle is determined by one of the area eigenvalues of the hole. The selected value is usually taken to be as close as possible to the one favored by the semiclassical quantum process. Alternatively, one can just calculate the transition probabilities between black-hole states by weighting them by exp$(S_2-S_1)$ where $S_1$ and $S_2$ are the entropies associated to the initial and final states. What is a black hole in LQG (see,  {\it e.g.}, \cite{Engle:2009vc,Krasnov:2009pd,Ashtekar:1998sp,FernandoBarbero:2009ai,Ghosh:2011fc}) ? It is basically an isolated horizon punctured by the edges of a spin network, that is a graph with edges labeled by $SU(2)$ representations and nodes characterized by intertwiners. An edge with spin representation $j$ carries an area of eigenvalue 
\begin{equation}
A_j=8\pi \gamma l_{Pl}^2\sqrt{j(j+1)},
\label{eq1}
\end{equation}
where $j$ is a half-integer and $\gamma$ is the Barbero-Immirzi parameter. A surface punctured by $N$ edges has a spectrum given by 
\begin{equation}
A_j=8 \pi \gamma l_{Pl}^2\sum_{n=1}^N\sqrt{j_n(j_n+1)},
\end{equation}
where the sum is carried out over all intersections of the edges with the surface. Each state with spin $j$ has a degeneracy $(2j+1)$. We believe that there might be two problems with the usual view of the Hawking evaporation in this framework. When one considers the transition from a state with mass $M_1$ to a state with mass $M_2$ which is, in general, very close to $M_1$ if the black hole is macroscopic, the actual final quantum state is completely different from the initial one most of the time. Even if the masses typically differ by much less than the Planck mass (and the areas differ by approximately the Planck area), the second quantum state corresponds to values of the spins 
that are in general completely different from the initial state. Using the quasidense distribution of states requires a complete reassigning of the quantum numbers to each puncture for every single  transition. This is in tension with a quantum gravitational origin of the evaporation process itself. If one considers the evaporation as due to a change of state of a given ``elementary area cell" (or, more precisely, to the settling down of the BH following this transition), there is no reason for all of the other elementary surfaces to change their quantum state at the same time. This was considered in, for example, \cite{Makela:2011vd}.
In addition, this raises an obvious second problem about causality: how can a ``far away" elementary cell know the way it has to change to adjust to the others? Our hypothesis here is that each particle evaporated by a black hole is basically due to the relaxation of the black hole following a change of state of a single elementary cell. We call this a local quantum gravity process and investigate it in the well-developed framework of LQG. In principle, however, the idea is quite general. 
It should be made clear that our hypothesis is nothing more than a reasonable alternative view that deserves to be studied until we have a fully dynamical quantum gravity description of the process. Ideally, one would build a model in which loop quantum gravity is coupled directly to quantum electrodynamics and compute the actual changes in the gravitational state upon emission of Hawking radiation. This is obviously beyond the scope of this study.  The key point here is that a purely local change of state of an ``elementary cell", with a small or moderate change in quantum number, is the  fundamental quantum gravity process taking place. This does {\it not} inconsistently assume that local physics knows the global BH quantities like temperature and mass. Once the quantum transition has taken place, without any {\it a priori} knowledge of the global picture, the BH relaxes through semiclassical processes according to the energy made available by the quantum transition. This automatically leads to a spectrum whose main classical properties agree with the Hawking evaporation. Of course, in the future, it would be important to investigate the settling down in a fully consistent way, going beyond the isolated horizon considered here which is, by construction, stationary.


\section{Low energy case}
Let us first consider the simpler case of a low-energy evaporation signal. It is easier for two reasons: first because the dynamics can then be neglected and, second, because there is no secondary emission associated with decaying particles in the sense that the black hole does not emit hadrons leading to gamma-rays. 

The main image is very simple and relies on the fact that the structure of a Schwarzshild black hole is such that (in Planck units) $dA=32\pi MdM$. If a quantum of energy $E\sim T$, with $T=1/(8 \pi M)$, is emitted,
it induces a BH mass change $dM=E$. The area change will then be $dA\sim4$. This is the main interest of the Bekenstein-Mukhanov hypothesis: assuming a discrete area spectrum with regularly spaced eigenvalues induces a line structure in the energies of the evaporation spectrum, even arbitrarily far away from the Planck mass. The very same change of area $dA$ (of order of the Planck area) will indeed lead to a relative line separation in the spectrum $dE/T$ which does {\it not} depend on the mass. \\

Following our hypothesis that the evaporation is due to a local change of state of a quantum of area, the spectrum will still exhibit lines in LQG. However, the eigenvalues of the area operator given by Eq. (\ref{eq1}) are a bit more subtle. In the large-$j$ limit, one recovers a regular line spectrum but the first eigenvalues are not equally spaced. 
If the resulting line spectrum is to have, as expected, the Hawking spectrum as an envelope, the favored transition is always one between two states $A_j$ and $A_{j-n}$ with $n$ a half-integer not much greater than unity: 95\% of the transitions will have $n\leq 3/2$. 

\begin{figure}[H]
\includegraphics[width=85mm,center]{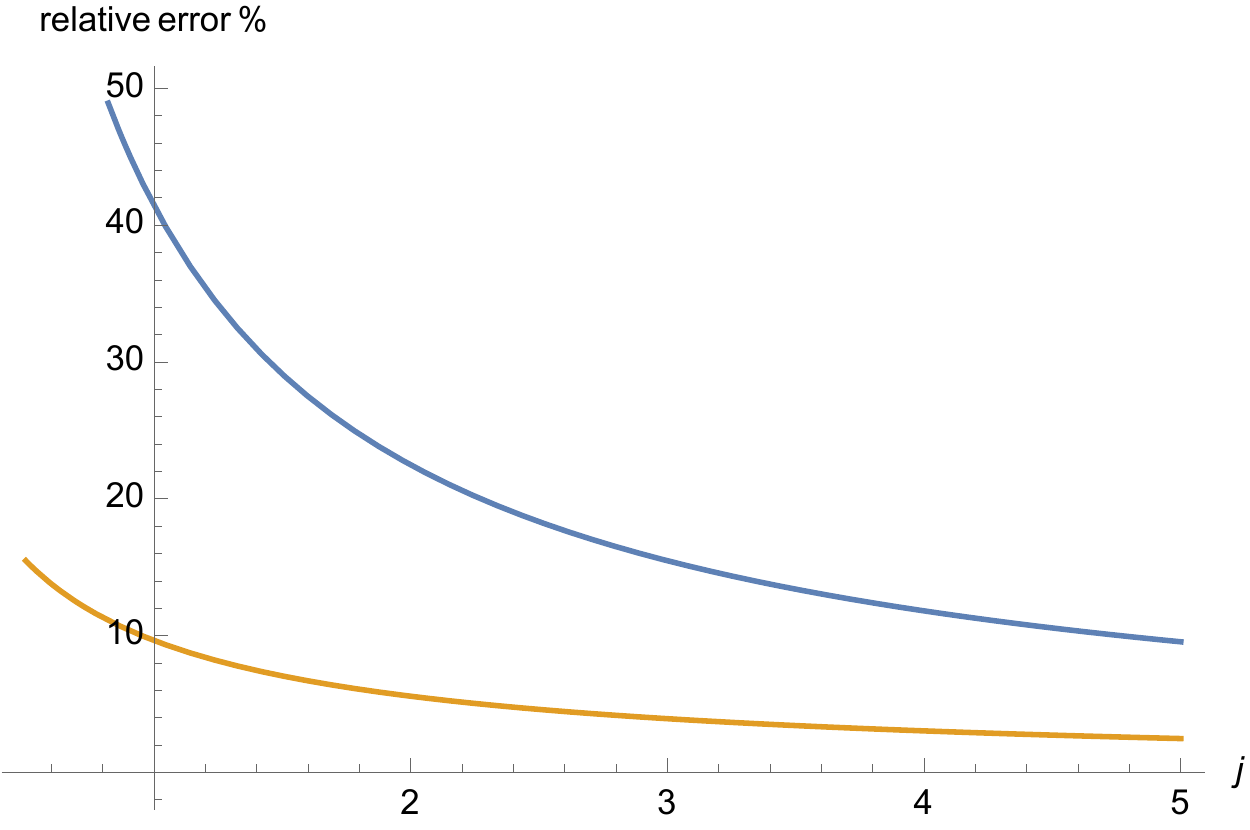}
\caption{Relative difference in the emitted energy between a purely regular line structure and the actual LQG line structure (in the local point of view of this study) in the $A_j\rightarrow A_0$ transitions (upper curve) and $A_j\rightarrow A_{1/2}$ transitions (lower curve).}
\label{error}
\end{figure}

The first studies of black holes in LQG (see, {\it e.g.}, \cite{Ashtekar:1997yu}) claimed that the punctures are mostly ``low-spin" ones. A  black hole 
is then expected to have most of his $j$'s close to 0. Transitions do exhibit generically a regular line structure. However, in some cases, largely corresponding either to transitions between $A_{j}$ and $A_{0}$ or between $A_j$ and $A_{1/2}$, there will be a deviation with respect to what would be expected from a regular line structure. This is what is shown in Fig. \ref{error}, which displays the relative energy difference between some $A_j\rightarrow A_0$ and $A_j\rightarrow A_{1/2}$  transitions and what would be expected from the same transitions in the regular Bekenstein-Mukhanov spectrum. The other way round, new models of holographic black holes were developed (see, {\it e.g.}, \cite{Ghosh:2013iwa}). Here, one uses the qualitative behavior of matter degeneracy suggested by standard QFT with a cutoff at the vicinity of the horizon -- {\it i.e.}, an exponential growth of vacuum entanglement in terms of the BH area. In this case, large $j$'s should dominate and the prediction is clearly that the line structure will be nearly perfect, as $A_{j}-A_{j-n}$ is very close to $n$ as soon as $j$ is much greater than unity. This opens up a very interesting possibility: not only should this effect allow to observe quantum gravity features at high masses, but it should also allow us to distinguish between  BH models in LQG.\\

An important question arises. Obviously, nearly equally spaced area eigenvalues and regular jumps between those values do {\it not} lead to the emission of quanta at the same energy as long as the evaporation goes on. When the black-hole area decreases by the same amount $dA$, the emitted energy varies like $1/M$. So, one should ensure that the change of area during the evaporation does not destroy the very possibility of observing lines. Let us evaluate the energy shift between two successive emissions associated with an identical area variation. Let us call $pA_0$ the area variation induced by the transition where $p$ is a half-integer and $A_0$ is the fundamental area of order $A_{Pl}$. It is easy to show that the relative variation of energy of the emitted quanta between consecutive emissions is, at lower order,
\begin{equation}
\frac{\Delta E}{E}\approx\frac{p}{2}\frac{A_0}{A}.
\end{equation}
This ensures that as long as the area is much larger than the Planck area, which is definitely the case for macroscopic black holes, the change in energy is negligible and the line structure can be observed if it exists: the fact that the BH mass changes during its evaporation does not wash out this interesting feature.\\

In practice, however, evaporating black holes are low-mass black holes, at  least when compared to a solar mass. This means that they must be primordial black holes (PBHs), except in some exotic low-Planck-scale models where they could be formed by collisions of particles in the contemporary Universe \cite{Barrau:2005zb}. The modes of production of PBHs are hypothetical (see, {\it e.g.}, \cite{Carr:1975qj,Carr:2009jm}). Various mechanisms have been considered. If we were to observe a single evaporating black hole, we would not care about its origin as far as the phenomenon studied here is concerned. Let us estimate the maximum distance at which this can be efficiently measured. The lifetime of the black hole is of order $M^3$, where $M$ is its initial mass. It is mostly determined by the emission of quanta of energy $E\sim T\sim 1/M$. There are, therefore, roughly $M^2$ quanta emitted in a time $M^3$, which means that the mean time between two emissions is of order $M$. The percentage of emitted quanta reaching a detector of surface $S$ is of order $S/R^2$ if the PBH is at distance $R$. The correct criterion for detection and identification of the signal consists of requiring a mean time $\Delta t$ between two measured photons from the same PBH to be smaller that a reference interval $\Delta t_0$ (otherwise the signal is lost in the background). This leads to the requirement $MR^2/S<\Delta t_0$, that is, a maximum detectable distance of
\begin{equation}
R_{max}\approx\sqrt{\frac{S\Delta t_0}{M}}.
\end{equation}
The most interesting case is, however, the signal emitted by a distribution of PBHs, whose masses are necessarily not exactly the same. Does the global line structure remains? This is not obvious, as different masses will induce different line energies and this might make the phenomenon experimentally invisible. By studying the energy of an emitted quantum in a given transition for two different BH masses, one can see that the relative energy variation is actually given by $\Delta M/M$. It is a quite general prediction of PBH formation mechanisms that their initial mass is roughly equal to the cosmological horizon mass at the formation time, $M\sim M_H \propto t$. As $t\propto T^{-2}$, where $T$ is the temperature of the Universe, 
\begin{equation}
\left| \frac{\Delta E}{E} \right|=2\left| \frac{\Delta T}{T} \right|.
\end{equation}
So if the relative change is to remain smaller than, say, 10\%, it is enough that the relative change in temperature during the formation period remains smaller than 5\%. This is reasonable for a PBH production associated, for example, with a phase transition (see, {\it e.g.} \cite{Cline:1996mk,Jedamzik:1999am}). 

\section{High energy case}

\begin{figure}[H]
\includegraphics[width=85mm,center]{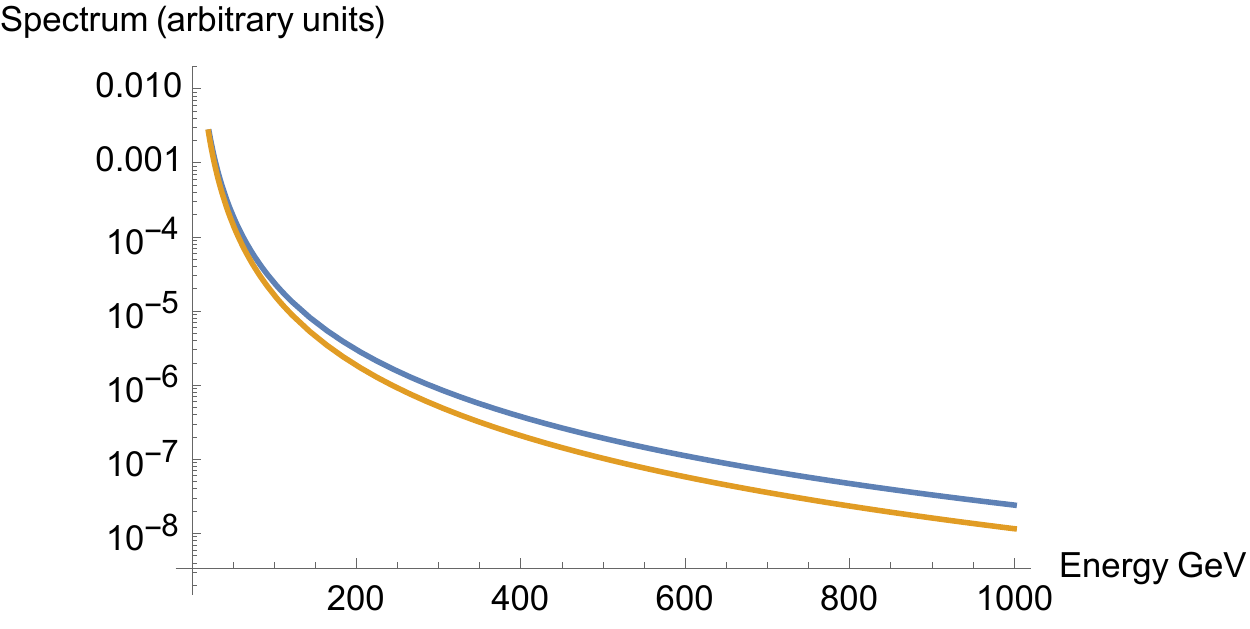}
\caption{Time integrated primary (upper curve) and secondary (lower curve) gamma-ray emission from an evaporating BH.}
\label{time}
\end{figure}

When the temperature of the black hole is greater than the QCD confinement scale (but still not much above the energies probed by accelerators), we conservatively assume that quarks are emitted and fragmentate into subsequent hadrons. This is a purely semiclassical prediction that does not rely on the underlying theory of quantum gravity. Some of those hadrons are unstable and will eventually decay into gamma-rays \cite{MacGibbon:1990zk,MacGibbon:1991tj}. Most gamma-rays emitted in this way come from the decay of neutral pions. We call this the secondary component. The problem is that even if the quarks are emitted with a spectrum made of lines, the resulting gamma-rays will obviously be distributed according to a continuum and the previously mentioned approach might not hold any longer. In addition, the instantaneous spectrum emitted by a black hole at a given temperature contains many more photons du to the secondary component than associated with the primary component (direct emission). We have investigated this point into the details by using the ``Lund Monte Carlo" PYTHIA code (with some scaling approximations in the low-energy range) \cite{Sjostrand:2014zea} to determine the normalized differential fragmentation functions $dg(Q,E)/dE$, where $Q$ is the quark energy and $E$ is the photon energy. It takes into account a large number of physics aspects, including hard and soft interactions, parton distributions, initial- and final-state parton showers, multiple interactions, fragmentation and decay. With this tool, we derived an analytical fit for the resulting fragmentation functions which describe the number of gamma rays generated between $E$ and $E+dE$ by the decay of the hadronization product of a quark with energy between $Q$ and $Q+dQ$. The secondary spectrum of gamma-rays reads as
\begin{eqnarray}
      \frac{d^2N_{\gamma}}{dEdt}&=&
      \sum_j\int_{Q=E}^{\infty}\alpha_j\Gamma_j(Q,T)
      \left(e^{\frac{Q}{T}}-(-1)^{2s_j}\right)^{-1}\\
      &\times&\frac{dg_{j\gamma}(Q,E)}{dE}{dQ},
\end{eqnarray}
where $j$ is the type of quark and $s_j=1/2$. The time-integration of this spectrum can obtained by writing
\begin{equation}
\frac{dN_{\gamma}}{dE}=\int_{M_i}^{M_f}\frac{d^2N_{\gamma}}{dEdt}\frac{dt}{dM}dM.
\end{equation}
The primary component of the instantaneous spectrum is a quasi-Planckian law (either a continuous one for the usual case or as the envelope of the lines for the quantum gravity case) but the secondary spectrum is more complicated. Mostly due to the decay of neutral pions, it can be roughly approximated by a Cauchy distribution near its maximum, and then an $E^{-1}$ power law followed by an exponential cutoff around the initial quark energy. It is continuous even if the primary emission is discrete. The time-integrated signal associated with the primary emission can easily be analytically shown to lead to a differential spectrum scaling as $E^{-3}$. We have performed the numerical integration of the secondary component. The very interesting point is that, as shown in Fig. \ref{time}, the time-integrated signal is nearly the same for both components. This is quite unexpected as the physics involved depends on the details of nongravitational processes (subtleties of the hadronization, cross sections for decays into gamma-rays, etc.). At a given BH temperature -- and, therefore, at a given mass -- the number of secondary photons is much higher than the number of primary photons. However, the mean energy of the secondary component is much smaller than for the primary component. This means that the primary emission was peaked at this energy when the BH mass was higher and, due the dynamics of the process $(dM/dt\propto -M^{-2})$, it has spent a ``longer time" in this mass region (between $M$ and $M+dM$). Those phenomena compensate each other and the neat result is that both components have the same order of magnitude.

The important consequence of this calculation is that, even in this case, the interesting quantum gravity line structure remains, in principle, detectable. It is not ``diluted" in a huge continuous signal due to the secondary component. The secondary component only induces a reasonable ``self-background" and as soon as the detector resolution is better than the line spacing, the phenomenon is easy to identify, if it exists at all. The relative energy difference between lines can easily be shown to be $\Delta E/E \sim \pi \gamma \sim 1$. As a typical detector resolution is of the order of $10\%-20\%$, the structure is clearly visible, even when taking into account this secondary component. 

 \section{Conclusion}

A quantum-gravity  ``local" perspective on the horizon structure of black holes 
might lead to a new view of the Hawking process: the evaporation would then be associated with field quanta emitted by the settling down of the black hole after the transition of a single ``surface element" between two area eigenstates. This is a speculative hypothesis that requires a more detailed theoretical investigation. 
However, if correct, we have shown that this would lead to a line structure in the spectrum, even for masses arbitrarily larger than the Planck mass. This is not washed out by the fact that black holes might be formed over a nonvanishing interval of masses. It also remains during the dynamics of the process in the sense that the energy variation between consecutive emissions is very small when compared with the separation between lines. More importantly it also remains visible when the secondary component, associated with the decay of unstable hadrons, is also taken into account. Finally, beyond being a ``smoking gun" candidate probe for quantum gravity, this would open interesting perspectives to discriminate between detailed loop quantum gravity models: high-spin models have a perfectly regular line structure whereas low-spin models exhibit some deviations with respect to the ideal case.

 \section{Acknowledgments}

A.B. would like to thank Marrit Schutten for the PYTHIA fits.

\bibliography{refs}

\begin{thebibliography}{23}
\expandafter\ifx\csname natexlab\endcsname\relax\def\natexlab#1{#1}\fi
\expandafter\ifx\csname bibnamefont\endcsname\relax
  \def\bibnamefont#1{#1}\fi
\expandafter\ifx\csname bibfnamefont\endcsname\relax
  \def\bibfnamefont#1{#1}\fi
\expandafter\ifx\csname citenamefont\endcsname\relax
  \def\citenamefont#1{#1}\fi
\expandafter\ifx\csname url\endcsname\relax
  \def\url#1{\texttt{#1}}\fi
\expandafter\ifx\csname urlprefix\endcsname\relax\def\urlprefix{URL }\fi
\providecommand{\bibinfo}[2]{#2}
\providecommand{\eprint}[2][]{\url{#2}}

\bibitem[{\citenamefont{Hawking}(1975)}]{Hawking:1974sw}
\bibinfo{author}{\bibfnamefont{S.~W.} \bibnamefont{Hawking}},
  \bibinfo{journal}{Commun. Math. Phys.} \textbf{\bibinfo{volume}{43}},
  \bibinfo{pages}{199} (\bibinfo{year}{1975}), \bibinfo{note}{[,167(1975)]}.

\bibitem[{\citenamefont{Page}(1976)}]{Page:1976df}
\bibinfo{author}{\bibfnamefont{D.~N.} \bibnamefont{Page}},
  \bibinfo{journal}{Phys. Rev.} \textbf{\bibinfo{volume}{D13}},
  \bibinfo{pages}{198} (\bibinfo{year}{1976}).

\bibitem[{\citenamefont{Bekenstein and Mukhanov}(1995)}]{Bekenstein:1995ju}
\bibinfo{author}{\bibfnamefont{J.~D.} \bibnamefont{Bekenstein}}
  \bibnamefont{and} \bibinfo{author}{\bibfnamefont{V.~F.}
  \bibnamefont{Mukhanov}}, \bibinfo{journal}{Phys. Lett.}
  \textbf{\bibinfo{volume}{B360}}, \bibinfo{pages}{7} (\bibinfo{year}{1995}),
  \eprint{gr-qc/9505012}.

\bibitem[{\citenamefont{Barreira et~al.}(1996)\citenamefont{Barreira, Carfora,
  and Rovelli}}]{Barreira:1996dt}
\bibinfo{author}{\bibfnamefont{M.}~\bibnamefont{Barreira}},
  \bibinfo{author}{\bibfnamefont{M.}~\bibnamefont{Carfora}}, \bibnamefont{and}
  \bibinfo{author}{\bibfnamefont{C.}~\bibnamefont{Rovelli}},
  \bibinfo{journal}{Gen. Rel. Grav.} \textbf{\bibinfo{volume}{28}},
  \bibinfo{pages}{1293} (\bibinfo{year}{1996}), \eprint{gr-qc/9603064}.

\bibitem[{\citenamefont{Yoon}(2016)}]{Yoon:2012cq}
\bibinfo{author}{\bibfnamefont{Y.}~\bibnamefont{Yoon}}, \bibinfo{journal}{J.
  Korean Phys. Soc.} \textbf{\bibinfo{volume}{68}}, \bibinfo{pages}{730}
  (\bibinfo{year}{2016}), \eprint{1210.8355}.

\bibitem[{\citenamefont{Barrau et~al.}(2011)\citenamefont{Barrau, Cailleteau,
  Cao, Diaz-Polo, and Grain}}]{Barrau:2011md}
\bibinfo{author}{\bibfnamefont{A.}~\bibnamefont{Barrau}},
  \bibinfo{author}{\bibfnamefont{T.}~\bibnamefont{Cailleteau}},
  \bibinfo{author}{\bibfnamefont{X.}~\bibnamefont{Cao}},
  \bibinfo{author}{\bibfnamefont{J.}~\bibnamefont{Diaz-Polo}},
  \bibnamefont{and} \bibinfo{author}{\bibfnamefont{J.}~\bibnamefont{Grain}},
  \bibinfo{journal}{Phys. Rev. Lett.} \textbf{\bibinfo{volume}{107}},
  \bibinfo{pages}{251301} (\bibinfo{year}{2011}), \eprint{1109.4239}.

\bibitem[{\citenamefont{Barrau et~al.}(2015)\citenamefont{Barrau, Cao, Noui,
  and Perez}}]{Barrau:2015ana}
\bibinfo{author}{\bibfnamefont{A.}~\bibnamefont{Barrau}},
  \bibinfo{author}{\bibfnamefont{X.}~\bibnamefont{Cao}},
  \bibinfo{author}{\bibfnamefont{K.}~\bibnamefont{Noui}}, \bibnamefont{and}
  \bibinfo{author}{\bibfnamefont{A.}~\bibnamefont{Perez}},
  \bibinfo{journal}{Phys. Rev.} \textbf{\bibinfo{volume}{D92}},
  \bibinfo{pages}{124046} (\bibinfo{year}{2015}), \eprint{1504.05352}.

\bibitem[{\citenamefont{Engle et~al.}(2010)\citenamefont{Engle, Perez, and
  Noui}}]{Engle:2009vc}
\bibinfo{author}{\bibfnamefont{J.}~\bibnamefont{Engle}},
  \bibinfo{author}{\bibfnamefont{A.}~\bibnamefont{Perez}}, \bibnamefont{and}
  \bibinfo{author}{\bibfnamefont{K.}~\bibnamefont{Noui}},
  \bibinfo{journal}{Phys. Rev. Lett.} \textbf{\bibinfo{volume}{105}},
  \bibinfo{pages}{031302} (\bibinfo{year}{2010}), \eprint{0905.3168}.

\bibitem[{\citenamefont{Krasnov and Rovelli}(2009)}]{Krasnov:2009pd}
\bibinfo{author}{\bibfnamefont{K.}~\bibnamefont{Krasnov}} \bibnamefont{and}
  \bibinfo{author}{\bibfnamefont{C.}~\bibnamefont{Rovelli}},
  \bibinfo{journal}{Class. Quant. Grav.} \textbf{\bibinfo{volume}{26}},
  \bibinfo{pages}{245009} (\bibinfo{year}{2009}), \eprint{0905.4916}.

\bibitem[{\citenamefont{Ashtekar et~al.}(1999)\citenamefont{Ashtekar, Beetle,
  and Fairhurst}}]{Ashtekar:1998sp}
\bibinfo{author}{\bibfnamefont{A.}~\bibnamefont{Ashtekar}},
  \bibinfo{author}{\bibfnamefont{C.}~\bibnamefont{Beetle}}, \bibnamefont{and}
  \bibinfo{author}{\bibfnamefont{S.}~\bibnamefont{Fairhurst}},
  \bibinfo{journal}{Class. Quant. Grav.} \textbf{\bibinfo{volume}{16}},
  \bibinfo{pages}{L1} (\bibinfo{year}{1999}), \eprint{gr-qc/9812065}.

\bibitem[{\citenamefont{Fernando~Barbero
  et~al.}(2009)\citenamefont{Fernando~Barbero, Lewandowski, and
  Villasenor}}]{FernandoBarbero:2009ai}
\bibinfo{author}{\bibfnamefont{G.~J.} \bibnamefont{Fernando~Barbero}},
  \bibinfo{author}{\bibfnamefont{J.}~\bibnamefont{Lewandowski}},
  \bibnamefont{and} \bibinfo{author}{\bibfnamefont{E.~J.~S.}
  \bibnamefont{Villasenor}}, \bibinfo{journal}{Phys. Rev.}
  \textbf{\bibinfo{volume}{D80}}, \bibinfo{pages}{044016}
  (\bibinfo{year}{2009}), \eprint{0905.3465}.

\bibitem[{\citenamefont{Ghosh and Perez}(2011)}]{Ghosh:2011fc}
\bibinfo{author}{\bibfnamefont{A.}~\bibnamefont{Ghosh}} \bibnamefont{and}
  \bibinfo{author}{\bibfnamefont{A.}~\bibnamefont{Perez}},
  \bibinfo{journal}{Phys. Rev. Lett.} \textbf{\bibinfo{volume}{107}},
  \bibinfo{pages}{241301} (\bibinfo{year}{2011}), \bibinfo{note}{[Erratum:
  Phys. Rev. Lett.108,169901(2012)]}, \eprint{1107.1320}.

\bibitem[{\citenamefont{Makela}(2011)}]{Makela:2011vd}
\bibinfo{author}{\bibfnamefont{J.}~\bibnamefont{Makela}},
  \bibinfo{journal}{Entropy} \textbf{\bibinfo{volume}{13}},
  \bibinfo{pages}{1324} (\bibinfo{year}{2011}), \eprint{1107.3975}.

\bibitem[{\citenamefont{Ashtekar et~al.}(1998)\citenamefont{Ashtekar, Baez,
  Corichi, and Krasnov}}]{Ashtekar:1997yu}
\bibinfo{author}{\bibfnamefont{A.}~\bibnamefont{Ashtekar}},
  \bibinfo{author}{\bibfnamefont{J.}~\bibnamefont{Baez}},
  \bibinfo{author}{\bibfnamefont{A.}~\bibnamefont{Corichi}}, \bibnamefont{and}
  \bibinfo{author}{\bibfnamefont{K.}~\bibnamefont{Krasnov}},
  \bibinfo{journal}{Phys. Rev. Lett.} \textbf{\bibinfo{volume}{80}},
  \bibinfo{pages}{904} (\bibinfo{year}{1998}), \eprint{gr-qc/9710007}.

\bibitem[{\citenamefont{Ghosh et~al.}(2014)\citenamefont{Ghosh, Noui, and
  Perez}}]{Ghosh:2013iwa}
\bibinfo{author}{\bibfnamefont{A.}~\bibnamefont{Ghosh}},
  \bibinfo{author}{\bibfnamefont{K.}~\bibnamefont{Noui}}, \bibnamefont{and}
  \bibinfo{author}{\bibfnamefont{A.}~\bibnamefont{Perez}},
  \bibinfo{journal}{Phys. Rev.} \textbf{\bibinfo{volume}{D89}},
  \bibinfo{pages}{084069} (\bibinfo{year}{2014}), \eprint{1309.4563}.

\bibitem[{\citenamefont{Barrau et~al.}(2005)\citenamefont{Barrau, Feron, and
  Grain}}]{Barrau:2005zb}
\bibinfo{author}{\bibfnamefont{A.}~\bibnamefont{Barrau}},
  \bibinfo{author}{\bibfnamefont{C.}~\bibnamefont{Feron}}, \bibnamefont{and}
  \bibinfo{author}{\bibfnamefont{J.}~\bibnamefont{Grain}},
  \bibinfo{journal}{Astrophys. J.} \textbf{\bibinfo{volume}{630}},
  \bibinfo{pages}{1015} (\bibinfo{year}{2005}), \eprint{astro-ph/0505436}.

\bibitem[{\citenamefont{Carr}(1975)}]{Carr:1975qj}
\bibinfo{author}{\bibfnamefont{B.~J.} \bibnamefont{Carr}},
  \bibinfo{journal}{Astrophys. J.} \textbf{\bibinfo{volume}{201}},
  \bibinfo{pages}{1} (\bibinfo{year}{1975}).

\bibitem[{\citenamefont{Carr et~al.}(2010)\citenamefont{Carr, Kohri, Sendouda,
  and Yokoyama}}]{Carr:2009jm}
\bibinfo{author}{\bibfnamefont{B.~J.} \bibnamefont{Carr}},
  \bibinfo{author}{\bibfnamefont{K.}~\bibnamefont{Kohri}},
  \bibinfo{author}{\bibfnamefont{Y.}~\bibnamefont{Sendouda}}, \bibnamefont{and}
  \bibinfo{author}{\bibfnamefont{J.}~\bibnamefont{Yokoyama}},
  \bibinfo{journal}{Phys. Rev.} \textbf{\bibinfo{volume}{D81}},
  \bibinfo{pages}{104019} (\bibinfo{year}{2010}), \eprint{0912.5297}.

\bibitem[{\citenamefont{Cline}(1996)}]{Cline:1996mk}
\bibinfo{author}{\bibfnamefont{D.~B.} \bibnamefont{Cline}},
  \bibinfo{journal}{Nucl. Phys.} \textbf{\bibinfo{volume}{A610}},
  \bibinfo{pages}{500C} (\bibinfo{year}{1996}).

\bibitem[{\citenamefont{Jedamzik and Niemeyer}(1999)}]{Jedamzik:1999am}
\bibinfo{author}{\bibfnamefont{K.}~\bibnamefont{Jedamzik}} \bibnamefont{and}
  \bibinfo{author}{\bibfnamefont{J.~C.} \bibnamefont{Niemeyer}},
  \bibinfo{journal}{Phys. Rev.} \textbf{\bibinfo{volume}{D59}},
  \bibinfo{pages}{124014} (\bibinfo{year}{1999}), \eprint{astro-ph/9901293}.

\bibitem[{\citenamefont{MacGibbon and Webber}(1990)}]{MacGibbon:1990zk}
\bibinfo{author}{\bibfnamefont{J.~H.} \bibnamefont{MacGibbon}}
  \bibnamefont{and} \bibinfo{author}{\bibfnamefont{B.~R.}
  \bibnamefont{Webber}}, \bibinfo{journal}{Phys. Rev.}
  \textbf{\bibinfo{volume}{D41}}, \bibinfo{pages}{3052} (\bibinfo{year}{1990}).

\bibitem[{\citenamefont{MacGibbon}(1991)}]{MacGibbon:1991tj}
\bibinfo{author}{\bibfnamefont{J.~H.} \bibnamefont{MacGibbon}},
  \bibinfo{journal}{Phys. Rev.} \textbf{\bibinfo{volume}{D44}},
  \bibinfo{pages}{376} (\bibinfo{year}{1991}).

\bibitem[{\citenamefont{Sjöstrand et~al.}(2015)\citenamefont{Sjöstrand, Ask,
  Christiansen, Corke, Desai, Ilten, Mrenna, Prestel, Rasmussen, and
  Skands}}]{Sjostrand:2014zea}
\bibinfo{author}{\bibfnamefont{T.}~\bibnamefont{Sjöstrand}},
  \bibinfo{author}{\bibfnamefont{S.}~\bibnamefont{Ask}},
  \bibinfo{author}{\bibfnamefont{J.~R.} \bibnamefont{Christiansen}},
  \bibinfo{author}{\bibfnamefont{R.}~\bibnamefont{Corke}},
  \bibinfo{author}{\bibfnamefont{N.}~\bibnamefont{Desai}},
  \bibinfo{author}{\bibfnamefont{P.}~\bibnamefont{Ilten}},
  \bibinfo{author}{\bibfnamefont{S.}~\bibnamefont{Mrenna}},
  \bibinfo{author}{\bibfnamefont{S.}~\bibnamefont{Prestel}},
  \bibinfo{author}{\bibfnamefont{C.~O.} \bibnamefont{Rasmussen}},
  \bibnamefont{and} \bibinfo{author}{\bibfnamefont{P.~Z.}
  \bibnamefont{Skands}}, \bibinfo{journal}{Comput. Phys. Commun.}
  \textbf{\bibinfo{volume}{191}}, \bibinfo{pages}{159} (\bibinfo{year}{2015}),
  \eprint{1410.3012}.

\end{thebibliography}
 \end{document}